\pdfoutput=1
\documentclass{emulateapj}
\usepackage{natbib}
\usepackage{color}

\usepackage{amsmath,amstext}

\shorttitle{The Revolution Revolution: Explaining the CV Period Gap}
\shortauthors{C. Garraffo et al.}

\begin{document}

\title{The Magnetic Nature of the Cataclysmic Variable Period Gap}
\author{C. Garraffo\altaffilmark{1,2}, J.~J. Drake\altaffilmark{1}, J.~D. Alvarado-Gomez\altaffilmark{1}, S.~P. Moschou\altaffilmark{1}, and O. Cohen\altaffilmark{3}}
\email{cgaraffo@cfa.harvard.edu}
\altaffiltext{1}{Harvard-Smithsonian Center for Astrophysics, 60 Garden Street, Cambridge MA 02138}
\altaffiltext{2}{Institute for Applied Computational Science, Harvard University, Cambridge MA 02138}
\altaffiltext{3}{Lowell Center for Space Science and Technology, University of Massachusetts, Lowell MA 01854}

\begin{abstract}

One of the most important problems in the context of cataclysmic variables (CVs) is the lack of observations of systems with periods between 2 and  3.12 hours, known as the period gap.  The orbital evolution of CVs with periods shorter than those in the gap is dominated by gravitational radiation while for periods exceeding those of the gap it is dominated by magnetic braking of the secondary star.  Spruit \& Ritter (1983) showed that as periods approach 3 hours and secondary stars become fully convective a sharp decline in magnetic dynamo and braking efficiency would result in such a gap.  Recent X-ray observations finding coronal magnetic energy dissipation is similar in fully convective and partly radiative M dwarfs cast this theory into doubt. In this work, we use Zeeman-Doppler imaging observations culled from the literature to show that the complexity of the surface magnetic fields of rapidly rotating M dwarfs increases with decreasing rotation period.  Garraffo et al.\ (2018) have shown that the efficiency of angular momentum loss of cool stars declines strongly with increasing complexity of their surface magnetic field. We explore the idea of \citet{Taam.Spruit:89} that magnetic complexity might then explain the period gap. By generating synthetic CV populations using a schematic CV evolutionary approach, we show that the CV period gap can naturally arise as a consequence of a rise in secondary star magnetic complexity near the long period edge of the gap that renders a sharp decline in their angular momentum loss rate. 

\end{abstract}


%@arxiver{fit.pdf, Mdot.pdf, R.pdf, Histogram.pdf, Visibility.pdf}

\section{Introduction}
\label{s:intro}

One of the most challenging puzzles in stellar evolution that emerged in the 1970s and early 1980s was the {\it cataclysmic variable period gap}.  Cataclysmic variables (CVs) are interacting binary stars comprising a white dwarf accreting from either a main-sequence or slightly evolved star, or from a brown dwarf.  As the population of known CVs grew, it became clear that systems with periods between 2 and  3.12 hours were rarely observed  compared with objects with shorter and longer periods \citep{Robinson:83,Ritter:84, Knigge.etal:11}.  While some of the underlying physics giving rise to this period gap are still debated, current models invoke changing rates of angular momentum loss as secondary stars are whittled down to lower and lower masses by attritional accretion onto their compact companion.

The orbital evolution of CVs with periods shorter than those in the gap is dominated by gravitational radiation \citep{Faulkner:71, Paczynski.Sienkiewicz:81}, while for periods exceeding those of the gap it is dominated by magnetic braking \citep{Eggleton:76, Verbunt.Zwaan:81}. In the latter regime, systems lose angular momentum via the magnetized winds of the non-degenerate companion and, as a consequence, their orbital separation is reduced and they spin up.  With mass from the secondary star being lost to the primary,  the secondary star drifts to later and later spectral type. By the time the system reaches the upper boundary of the period gap ($\sim 3.12$ hours),  the secondary star has been reduced to the mass of a fully-convective M dwarf  (\citealt{Robinson.etal:81,Spruit.Ritter:83}, and references therein). 

\cite{Spruit.Ritter:83} and \citet{Rappaport.etal:83} showed that a fast decrease in angular momentum loss of $\sim 90\%$ as the secondary approaches the fully-convective limit would result in strong suppression of the mass accretion from the secondary star onto its companion that would explain the appearance of the period gap.  A detailed review of this evolutionary scenario has been provided by \citet{Knigge.etal:11}. 
Typically, CVs are discovered through UV or X-ray emission from heated accreting material: if the accretion stops the systems cannot be easily discerned.  

The underlying motivation for a fairly abrupt angular momentum loss reduction at the fully convective limit stems from arguments that magnetic dynamo action in Sun-like stars occurs at the interface between the convection zone and the radiative interior---the ``tachocline''; \citep[see, e.g.,][]{Spruit.Ritter:83}.  
The fact that the secondary star becomes fully convective near the edge of the period gap led to the idea that there is a fundamental change in the effectiveness of the dynamo at this limit. This supposed demise of the dynamo thus results in a magnetic braking ``disruption''. 
However, evidence for such a dynamo demise has historically been weak or lacking. Recently, \cite{Wright.Drake:16} have found that the X-ray emission level as a function of stellar rotation period is essentially the same for both fully convective, and partly convective and more Sun-like stars for both rapid and slow rotators.  X-ray activity has been shown to be a good proxy for surface magnetic flux \citep[e.g.][]{Pevtsov.etal:03}. \citet{Wright.Drake:16} then argued that the invariance of X-ray behavior regardless of the presence or absence of a radiative zone supports the arguments advanced by \citet{Spruit:11} that magnetic dynamo action is instead distributed in the convection zone.  The collateral effect of the \citet{Wright.Drake:16} results is that, at face value, there is no change in the surface magnetic activity that drives magnetic braking: the {\it disrupted magnetic braking} theory for the CV period gap is broken.  \citet{Taam.Spruit:89} had anticipated this and suggested that the magnetic field of the secondary star might grow in complexity at the edge of the period gap, reducing the number of open field lines and the subsequent angular momentum loss.

Recently, it has been shown using different types of magnetohydrodynamic (MHD) wind models that the efficiency of angular momentum loss of cool stars strongly depends on the complexity of their stellar surface magnetic fields \citep[][from hereon CG16]{Reville.etal:15a, Garraffo.etal:15, Garraffo.etal:16a}, echoing the earlier analytical conclusions of \citet{Taam.Spruit:89}.  \citet[][hereafter CG18]{Garraffo.etal:18} provided a new predictive spin-down model based on sophistcated MHD wind modeling that accounts for this magnetic modulation, and assumes that the complexity of the magnetic field is a function of Rossby number {\it Ro} ({\it Ro} $ = P_{rot}/\tau$, where $\tau$ is convective turnover time).  This assumption is well supported by Zeeman-Doppler imaging (ZDI) observations of the magnetic fields on the surfaces of Sun-like stars showing that faster rotating stars store a larger fraction of their magnetic flux in higher order multipole components of the field \citep[e.g.,][]{Donati:03, Donati.Landstreet:09, Marsden.etal:11a, Waite.etal:11, Waite.etal:15}. As a consequence, they lose angular momentum much less efficiently. 

Magnetic braking of CVs is usually modeled considering the secondary star as a single star and assuming it has a simple, fixed magnetic configuration such as a dipole. 
In this work, we study the available ZDI observations of late M-dwarfs by \cite{Morin.etal:10} to infer the underlying geometry of their magnetic fields as a function of rotation period.  We follow the idea of \citet{Taam.Spruit:89} and explore the possibility that there is an disruption of angular momentum loss near the upper boundary of the period gap ($\sim 3.12$) resulting from an increasing magnetic complexity of the secondary star.  %We perform spherical harmonics decomposition for our sample of late M-dwarfs magnetic maps, compute 
Using the observed magnetic geometry of M Dwarfs as a function of rotation period derived here and the spin-down model provided in CG18,  we use the schematic method of \citet{Spruit.Ritter:83} to reconstruct the orbital evolution of CVs near the gap period. Using this evolutionary recipe and the expected formation rate of CVs, we generate synthetic populations and predict the fraction of systems and their visibility as a function of orbital period.  

%and find that, indeed, an increase in complexity near the upper boundary of the period gap results in the lack of observable systems in a range. %We study what is the complexity as a function of period that best reproduces the period gap range.
%We find that the complexity function that best explains the period gap is consistent with the trend shown by all the available observations of stellar magnetic fields in late M dwarfs.  
%This provides both a simple and observed physical basis for the period gap, and a more realistic orbital evolution model for CVs above and within the period gap. 

The paper is organized as follows. In Section~\ref{mdwarfs} we compute the complexity of the available ZDI observations for late M Dwarfs and compare our data to the complexity function assumed in CG18. In Section~\ref{aml} we use that function and the prescription for angular momentum loss from CG18 to describe a single system's orbital evolution. In Section~\ref{ps} we generate and evolve synthetic populations and compare them with observations. In Section~\ref{results} we discuss our findings and summarize our main conclusions. 

\section{Late M Dwarf Magnetic Complexity}
\label{mdwarfs}

The ZDI technique enables inference of the large scale magnetic morphology of active cool stars fairly robustly given that the phase coverage of the observations is complete enough, or sufficiently high signal-to-noise ratios are achieved in poorly sampled datasets \citep{Donati.Brown:97,Hussain.etal:00,Alvarado-Gomez.etal:15}.   
There is growing evidence from ZDI observations that faster rotating Sun-like stars show a more complex magnetic morphology on their surface \citep[e.\ g.][]{Donati:03, Donati.Landstreet:09, Marsden.etal:11, Waite.etal:15, Alvarado-Gomez.etal:15}.   The importance of these observations lies in the results of 
MHD models of solar-like stellar winds that confirm the idea of \citet{Taam.Spruit:89} and predict that an increase in the complexity of the magnetic fields should lead to a strong suppression of angular momentum loss efficiency \citep{Reville.etal:15a, Garraffo.etal:15}.   It was shown by \cite{Garraffo.etal:13, Garraffo.etal:15} that only the first few terms ($Y^m_n(\theta,\psi) = N\, e^{im\psi} P^m_n(\cos{\theta})$, $n \lessapprox 7$) in a spherical harmonics decomposition of the surface field are relevant to this effect. At the same time, those first moments are the ones for which the ZDI technique is most reliable. 
%Therefore, we can use ZDI maps to infer the evolution of the large scale magnetic field's complexity as a function of rotation period that is is important for their orbital evolution.   
We examine this here for late M dwarfs and study how it affects their rotational evolution.  
%We do so through a systematic study of observed M dwarfs magnetic field geometry as a function of rotation period.

\begin{figure}{}
\center
\includegraphics[trim = .35in 0.2in
  0.in .2in,clip, width =  0.45\textwidth]{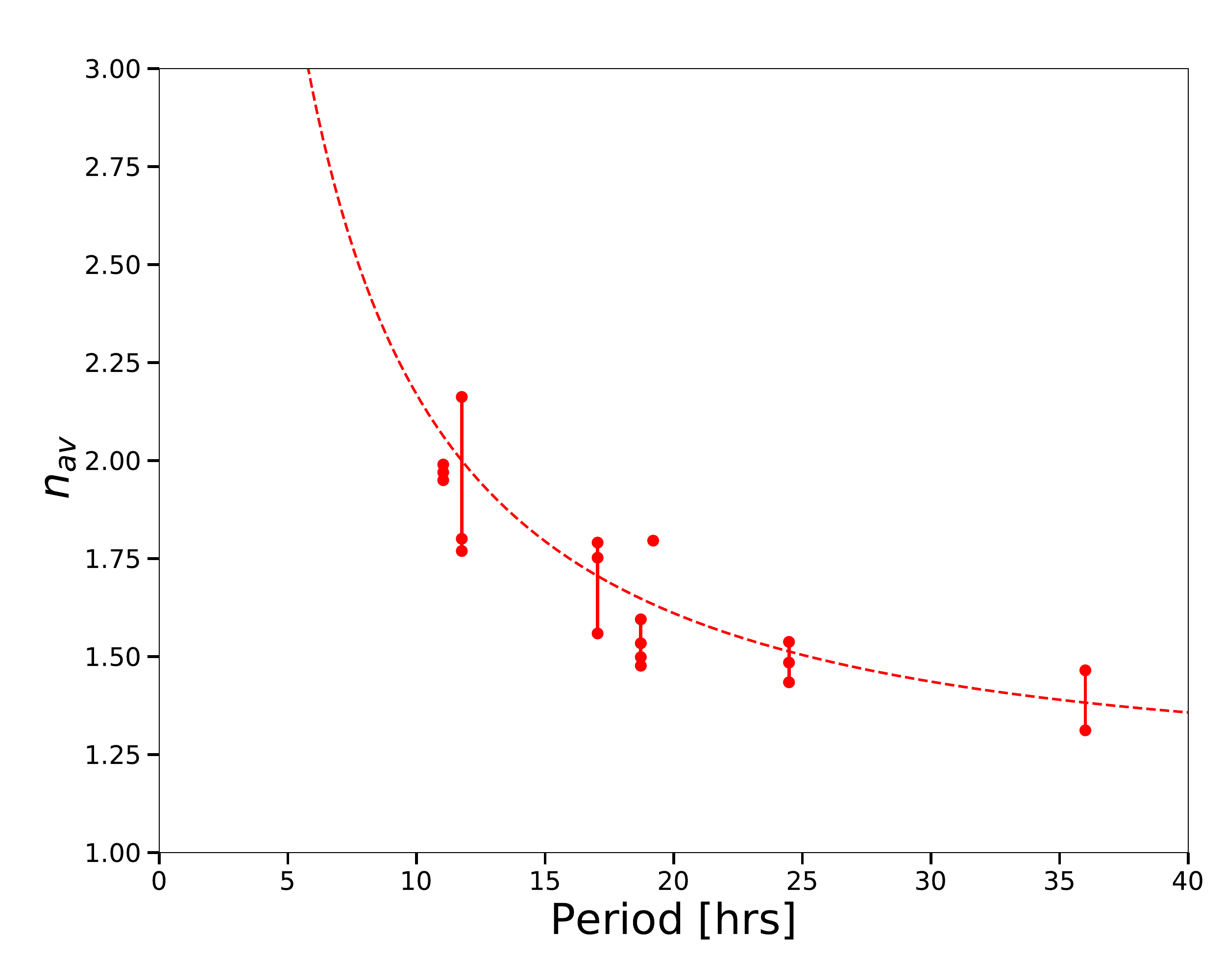} 
\caption{The magnetic complexity of available late M dwarf ZDI observations (dots) and the complexity function relating complexity, $n_{av}$, and rotation period from CG18 for a convective turnover time of $\tau_c=100$~days (dashed curve). The connected dots represent different ZDI observations of the same star.\\
}
\label{fig:mds}
\end{figure}

We compute the large scale magnetic complexity for all 19 available radial magnetic maps\footnote{GJ~51 (2006, 2007, and 2008), GJ~1156 (2007, 2008, and 2009), GJ~1245 (2006, 2007, and 2008), WX~UMa (2006, 2007, 2008, and 2009), DX~Cnc (2007, 2008, and 2009), GJ~3622 (2008, 2009), VB~10 (2009).} for the seven late M dwarfs (M5--M8) observed by \cite{Morin.etal:10} following the spherical harmonic decomposition method of \cite{Garraffo.etal:16a}.  In this approach, the representative average complexity, $n_{av}$, is given by 
\begin{equation}
\, \, \, n_{av} = \sum_{n=0}^{n_{max}} \frac{n \, F_n}{F_T},
\label{eq:nav}
\end{equation}
where $F_n$ is the magnetic flux in each $n$-order term in the spherical harmonic decomposition and $F_T$ is the total flux in the original magnetogram.   CG16 and \cite{Finley.etal:18} showed that angular momentum loss rates are independent of the different azimuthal distributions of magnetic flux ($m$ modes) for a given complexity ($n$ mode). It is on this basis that we neglect the parameter $m$ in the decomposition.  The results of application of Equation~\ref{eq:nav} for the \cite{Morin.etal:10} late M dwarfs are plotted as a function of stellar rotation period in Figure~\ref{fig:mds}.  The ZDI maps show a trend of increasing complexity of the surface magnetic field for faster rotating stars that becomes more pronounced at the shortest periods of a few hours.  

{}

%\textcolor{blue}{
We compare the complexity derived from the observations to the magnetic complexity function in CG18, $n = \frac{a}{Ro} +Ro+1$ in Figure~\ref{fig:mds}. Here, $Ro$ is the commonly used {\em Rossby} number for stellar magnetic dynamo activity representing the ratio $Ro=\tau_c/P_{rot}$ of rotation period $P_{rot}$, and convective turnover time, $\tau_c$.  This complexity function was derived by matching the evolution of rotation periods of stars in young open clusters using the relationship between magnetic complexity and angular momentum loss derived by CG16. 

The function shown in Figure~\ref{fig:mds} that provides a good match to the data corresponds to a convective turnover time $\tau_c = 100$ and value $a=0.005$.  The former is
appropriate for the spectral types of the secondary stars considered here \citep{Morin.etal:10, Wright.etal:11}. 
%{\bf JJD: here for different value of a adopted?  Should we add the a=0.01 curve? (I'm a bit confused about the other factor of 2 in Jdip in you code)}
The value of $a$ is 4 times smaller than the one used by CG18, though we note that this quantity only affects the most rapidly rotating stars with the smallest Rossby numbers and was not well-constrained in that study that only dealt with stars with masses greater than $0.3 M_\odot$.  In our CV orbital evolution model (Section~\ref{aml}) we adopt an intermediate value $a=0.01$.
The shape of the complexity function, $n$, which dictates the change in angular momentum loss with changing rotation period, remains the same. 
It should also be borne in mind that, while $n_{av}$ was shown to work reasonably well for parameterizing real magnetograms \citep{Garraffo.etal:16a}, there is some arbitrarity in its definition and it should strictly be interpreted only as a relative measure of complexity.  The spatial resolution of ZDI-reconstructed magnetic fields is also limited by the rotation-induced velocity shift of surface magnetic features and the deconstructed complexity is inevitably going to be a lower limit to the true complexity.

%{\bf \textcolor{blue}{Now a=0.01, which is 1/2 of the value in The Revolution Revolution. This has nothing to do with the factor 2 (which we don't need to worry about, it is a constat value in Jdip that we don't mention). The plotted curve corresponds to a=0.005 that we are not using now. In summary, the model has a=0.02, the data fits with a=0.005, we use something in between a=0.01 that gives us approx the same Jdip that spruit for periods just above the gap. We can argue that the data fits a=0.005 instead of 0.01 because of a systematic loss of complexity in the ZDI technique. }}
%}

%that best reproduced the period gap (see Equation~\ref{nav}) and we see that the observations further support our assumption.

\section{Magnetic Disruption of Angular Momentum Loss}
\label{aml}

\subsection{Angular momentum loss}

%\cite{Spruit.Ritter:83} have shown that a sudden decrease of $\sim 90\%$ in angular momentum loss at the right rotation period would lead to a drastic reduction of the mass accretion and, thus, explain the lack of observations in the period gap.  We explore the possibility of this decrease being a result of increasing magnetic complexity. 
%It has been shown that the large scale magnetic morphology of the stellar surface magnetic fields can significantly change the efficiency of the angular momentum loss \citep{Reville.etal:15a, Garraffo.etal:15}. The orbital evolution of CVs is dominated by the magnetic braking of the late-type companion for systems with orbital periods over 3.2 hours.  The connection between magnetic complexity and magnetic braking efficiency established by \citet{Taam.Spruit:89}, \citet{Reville.etal:15a} and \citet{Garraffo.etal:15} means that the evolution of magnetic complexity of late M dwarfs with time will factor into the angular momentum evolution of systems on the long period side of the period gap.  
%An increasing number of ZDI observations suggest an faster rotating stars store a larger fraction of their magnetic flux in higher order components of the field.  
Here, we simulate the orbital evolution of CVs, including, for the first time, the magnetic complexity modulation and its effect on angular momentum loss.  We use the same spin-down model as presented by CG18.%\cite{Garraffo.etal:16}.  

As systems evolve and lose angular momentum, the orbits shrink and periods decrease.  Efficient spin-orbit coupling means that the secondary stars spin up, maintaining synchronicity between their rotation and orbital periods.  Figure~\ref{fig:mds} indicates that their magnetic complexity will also increase as they spin-up.  The data show a somewhat
steeper increase in complexity for the stars with the shortest rotation periods, although data are lacking for periods less than 10 hours.   We proceed by adopting the CG18 spin-down model and complexity function, 
taking Figure~\ref{fig:mds} as offering some empirical support (with the function reproducing well the data on average) but recognizing that it still represents somewhat of an extrapolation to reach the CV period gap.  The spin down model can then be described by the following equations (see CG18 for further details):
\begin{flalign}
&\dot{J} =  \dot{J}_{Dip} Q_{J}(n_{av}) &
\label{eq:Jdot} 
\end{flalign}
%??Remove eqn below??
%\begin{flalign}
%&\dot{J}_{Dip} = c \cdot \Omega^3 \tau_c&
%\label{eq:Jdip} 
%\end{flalign}
\begin{flalign}
&Q_{J}(n_{av}) = 4.05 \, e^{-1.4 n_{av}}+(n_{av}-1)/(60 B\, n_{av}) &
\label{eq:Q}
\end{flalign}
%??Make this consistent with removing eqn 3.  Also make sure Eqn 3 is not referred to elsewhere, and add the statement that there is an implicit assumption of constant magnetic field strength, which is justified by the observation that stars with Ro < 0.1 are in the saturated regime and their X-ray luminosities are invariant with rotation period.??
where $\dot{J}$ represents the angular momentum loss rate, $\dot{J}_{Dip}$ the rate for a dipolar magnetic configuration.
%, and $\tau_c$ the convective turnover time. 
The factor $Q_{J}(n_{av})$ includes the magnetic morphology dependence, $B$ stands for surface field strength [Gauss], and $n$ stands for the magnetic multipolar moment describing the complexity of the field.  As in CG18, we neglect the second term in Equation \ref{eq:Q} because it is negligible for $n_{av}<7$, which is the regime we are exploring here.  As noted in Section~\ref{mdwarfs}, we adopt a value $a=0.01$, which gives  $\dot{J}/J_{orb} \approx 1 \times 10^{-8}$ for a period of 5~hrs and 
$\approx 2. \times 10^{-9}$ just above the gap at 3.2 hrs period.  This can be compared with the constant value $\approx 5 \times 10^{-9}$ yr$^{-1}$ used by \citet{Spruit.Ritter:83} for all periods approaching and through the period gap.

\subsection{\bf Schematic analysis of system evolution}

%We assume the same complexity evolution with time as CG18, which is consistent with ZDI observations (see Figure~\ref{fig:mds}).  
The magnetic complexity increase of the secondary star as the system spins up towards the $3.2$ hours period is responsible for a decrease in angular momentum loss efficiency.  We show here that the decrease happens to be large and fast enough to suppress mass accretion and explain the period gap,  as predicted by \cite{Spruit.Ritter:83,Rappaport.etal:83,Taam.Spruit:89}.  

Following the approaches of \cite{Spruit.Ritter:83} and \cite{Knigge.etal:11}, we model the evolution of a single system as it approaches the upper boundary of the period gap. 
We emphasise that the method used here is schematic and based on homologous stellar models calibrated with the \cite{Grossman.etal:74} main sequence, and should ultimately be verified using more detailed stellar evolution models.
%like MESA \citep{Paxton.etal:11} (which we plan to do in the future). 
%The goal in this work is to show that the mechanism described by \cite{Spruit.Ritter:83} explains the period gap without the need of a suppression of the dynamo at the fully convective limit. 
%In our approach, the magnetic braking "interruption" is a result of the complexity evolution of the stellar surface magnetic fields rather than from a dynamo shut off.  
Such validation of the Spruit \& Ritter magnetic disruption mechanism for producing a period gap when using full stellar evolution models has recently been presented by \citet{Paxton.etal:15} and \citet{Kalomeni.etal:16}, and references therein.  
%We follow the same approach but with a magnetic braking "interruption" (also rapid and of $\approx 90 \%$ ) that arises naturally from the evolution of magnetic complexity.  
The difference in our study is that $\dot{J}_{dip}/J$ does not change instantaneously, but still changes abruptly (much faster than the Kelvin-helmholtz timescale) and by the same amplitude  of $\approx 90 \%$ as in \cite{Spruit.Ritter:83}, and near the upper boundary of the period gap. While this could potentially make a difference in a detailed stellar evolutionary model, the similarity in the angular momentum loss suppression and the fact that the homologous model results in detachment lends support for this proof of concept study.
%We do plan to run the MESA simulations as a next step. }%\citep[see, for example, ][]{Howell.etal:01, Kalomeni.etal:16}

We assume the secondary star has an initial mass of $M_2= 0.42 M_{\odot}$ and an initial radius of $ R_e \approx 0.9\, M^{0.8}$, which is that expected from the mass-radius relation for a lower-main sequence star in thermal equilibrium \citep{Whyte.Eggleton:80, Iben.Tutukov:84, Knigge.etal:11}.  If the accretion timescale ($\tau_{\dot{M_2}} = M_2/\dot{M_2} $, where $\dot{M_2}$ is the secondary's mass loss rate through accretion) is shorter than the thermal (Kelvin-Helmholtz) timescale of the donor star, $\tau_{kh} \sim G\, M^2/R$, then the secondary star is taken out of thermal equilibrium and inflates as a consequence of adiabatic mass loss. The mass-radius relation in the adiabatic limit becomes $R\propto M^{-1/3}$ \citep{Rappaport.etal:82}, but CV donors turn out to be somewhere in between thermal equilibrium and the adiabatic limit when accreting fast enough. Theoretical expectations and observations suggest that in this regime $R \sim M^{0.65}$ \citep{Patterson.etal:05, Knigge:06, Knigge.etal:11}.  When approaching the minimum period ($\approx 80$ min), the angular momentum loss through gravitational radiation increases rapidly, further pushing the secondary star towards the adiabatic regime and, as a consequence, the mass-radius relationship becomes $R \sim M_2^{0.21}$ \citep{Knigge.etal:11}.  We use this relationship for systems with $M_2 < 0.1 M_{\odot}$ minutes. 

%\textbf{ This angular momentum loss rates increase rapidly with shorter periods, further pushing the secondary star towards the adiabatic regime. Meanwhile, the secondary star has became less massive. In this regime ($M_2<0.1 M_{\odot}$), then, we use a mass-radius relationship $R_2 \propto M_2^{0.21}$ that reflects this close-to-adiabatic mass transfer \citep{Knigge.etal:11}. }    

We evolve the system for $10^9$ years with a time step of $10^5$ years. For each step we calculate the angular momentum loss using the equations above (CG18) with the complexity function discussed in Section \ref{mdwarfs}.  In addition, we include the angular momentum loss due to gravitational radiation given by \cite{Paczynski:67}, 
\begin{equation}\dot{J}_{GR} = - \frac{32}{5} \frac{G^{7/2}}{c^5} \frac{M^2_1 M^2_2 M^{1/2}}{a^{7/2}},
\end{equation}
with $G$ being the universal gravitational constant, $M = M_1+M_2$, $a$ the orbital separation, and $c$ the speed of light.  This contribution to the spin-up process becomes important for periods shorter than $\sim$ 2 hrs. 
 
Whenever the donor star is in contact with the Roche lobe there is mass transfer. We compute the accretion rate using \cite{Knigge.etal:11} equations (1)-(3) and assume $R \sim M^{0.65}$. If $\tau_{\dot{M_2}}<\tau_{kh}$, the donor will detach from its Roche lobe and its radius will decrease at a rate $\dot{R_2} \sim (R_2-R_e)/\tau_{kh}$ towards thermal equilibrium. The thermal timescale of the reference donor star used by \citet{Spruit.Ritter:83} above the period gap ($M=0.28 M_\odot$, $R=0.285R_\odot$) is $\tau_{kh} \approx 3 \times 10^8$.  We use this reference value to set the normalization factor for $\tau_{kh} \propto G\, M^2/R$.  It might be argued that the actual timescale for the radius to adjust to a change in accretion rate is shorter than the equilibrium Kelvin-Helmholtz scale $\tau_{kh}$ \citep[see, for example,][]{Stehle.etal:96, Knigge.etal:11}.  In our evolutionary model we find the exact value of the radius shrinkage timescale is not critical for detachment, as long as the change in $\dot{J}$ is fast enough to overcome the decrease in the orbital angular momentum so that the ratio $J_{orb}/\dot{J}$ increases.  This is satisfied in our model for both $\tau_{kh}$ and $\tau_{adj}$; $\dot{J}/J$ decreases on a timescale of approximately $2\times 10^7$~yr, to be compared with a value of $\tau_{adj}\approx 8\times 10^7$~yr at a period of 3~hr \citep{Knigge.etal:11}.

As the system evolves, the donor star is losing mass and becoming of later spectral type.  We account for this with a convective turnover time that evolves with the mass of the secondary star following \cite{Wright.etal:11}.  However, our results do not change qualitatively when assuming a constant convective turnover time consistent with those expected for late M dwarfs ($\sim 100$).  

At each step, the orbital period follows from 
\begin{equation}
P = 9 \pi (2\, G)^{-1/2} \left(\frac{R^3_2}{M_2}\right)^{1/2}
\label{period}
\end{equation}
\citep{Paczynski.Sienkiewicz:81, Spruit.Ritter:83}. 
%For the initial masses considered here, this period turns out to be $\sim 4.82$ hrs.  

\begin{figure}{}
\center
\includegraphics[trim = 1.2in 3.95in
  1.2in 3.2in,clip, width =  0.5\textwidth]{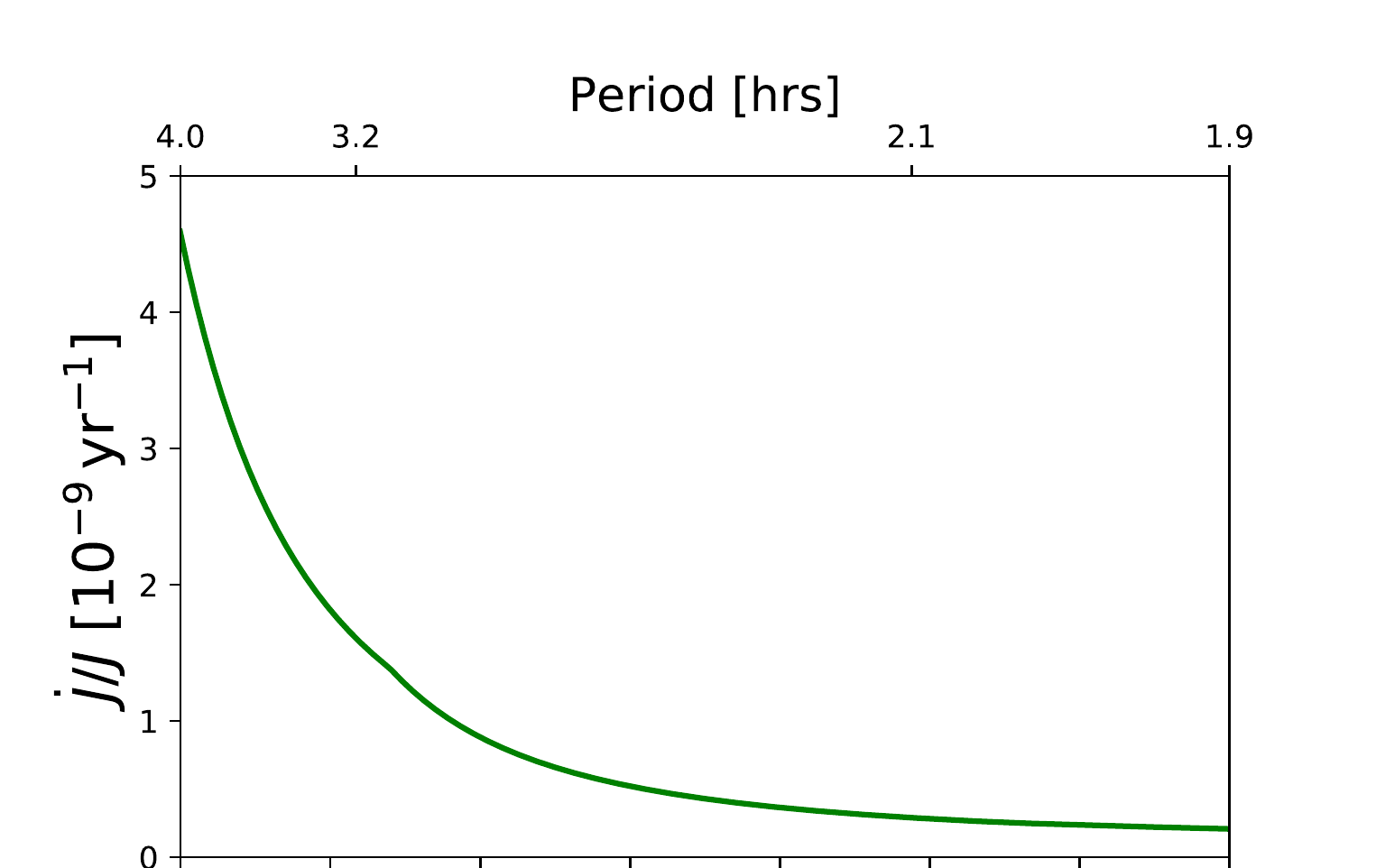}\\
   \includegraphics[trim = 1.2in 3.95in
  1.2in 3.9in,clip, width =  0.5\textwidth]{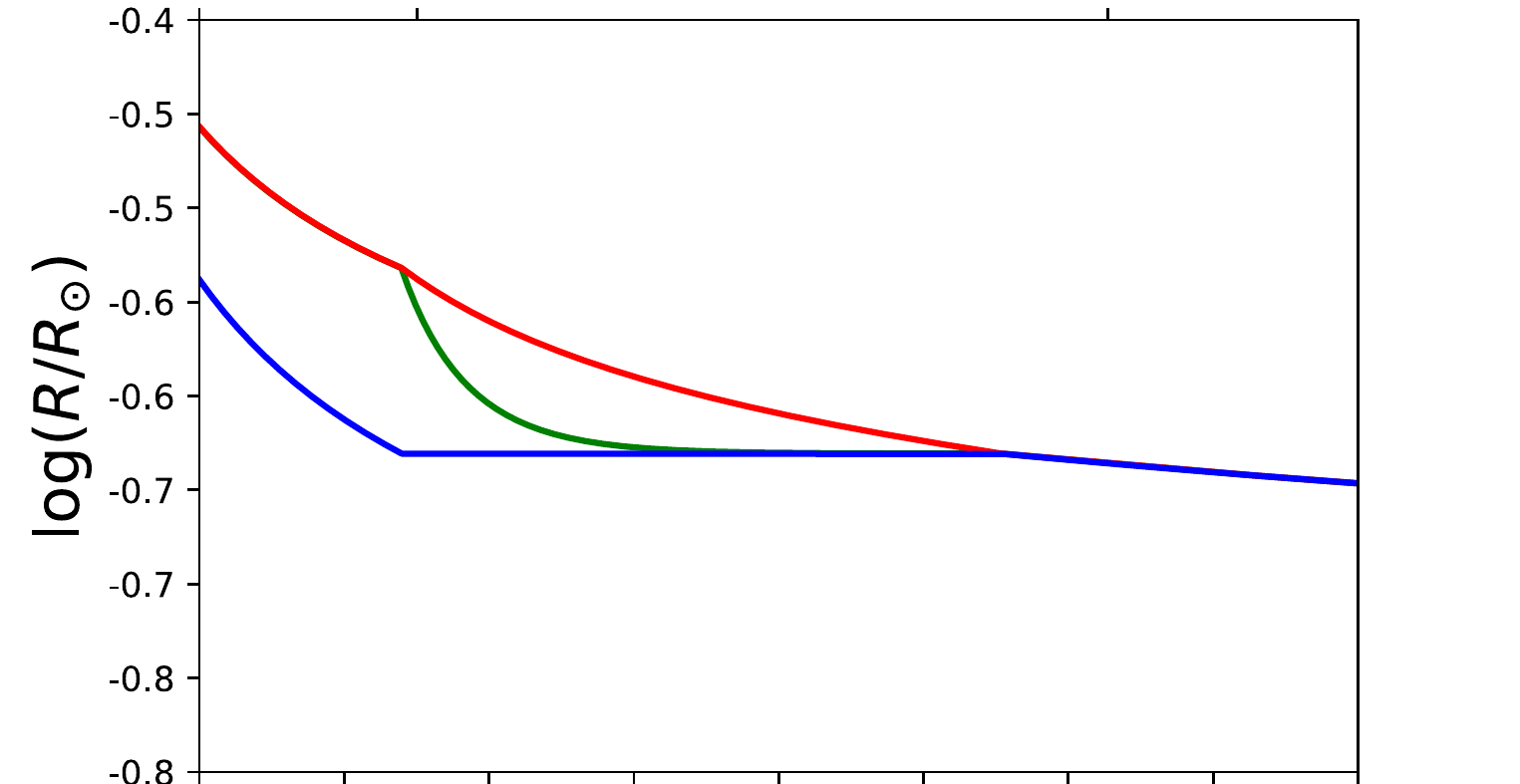}\\
   \includegraphics[trim = 1.2in 3.2in
  1.2in 3.9in,clip, width =  0.5\textwidth]{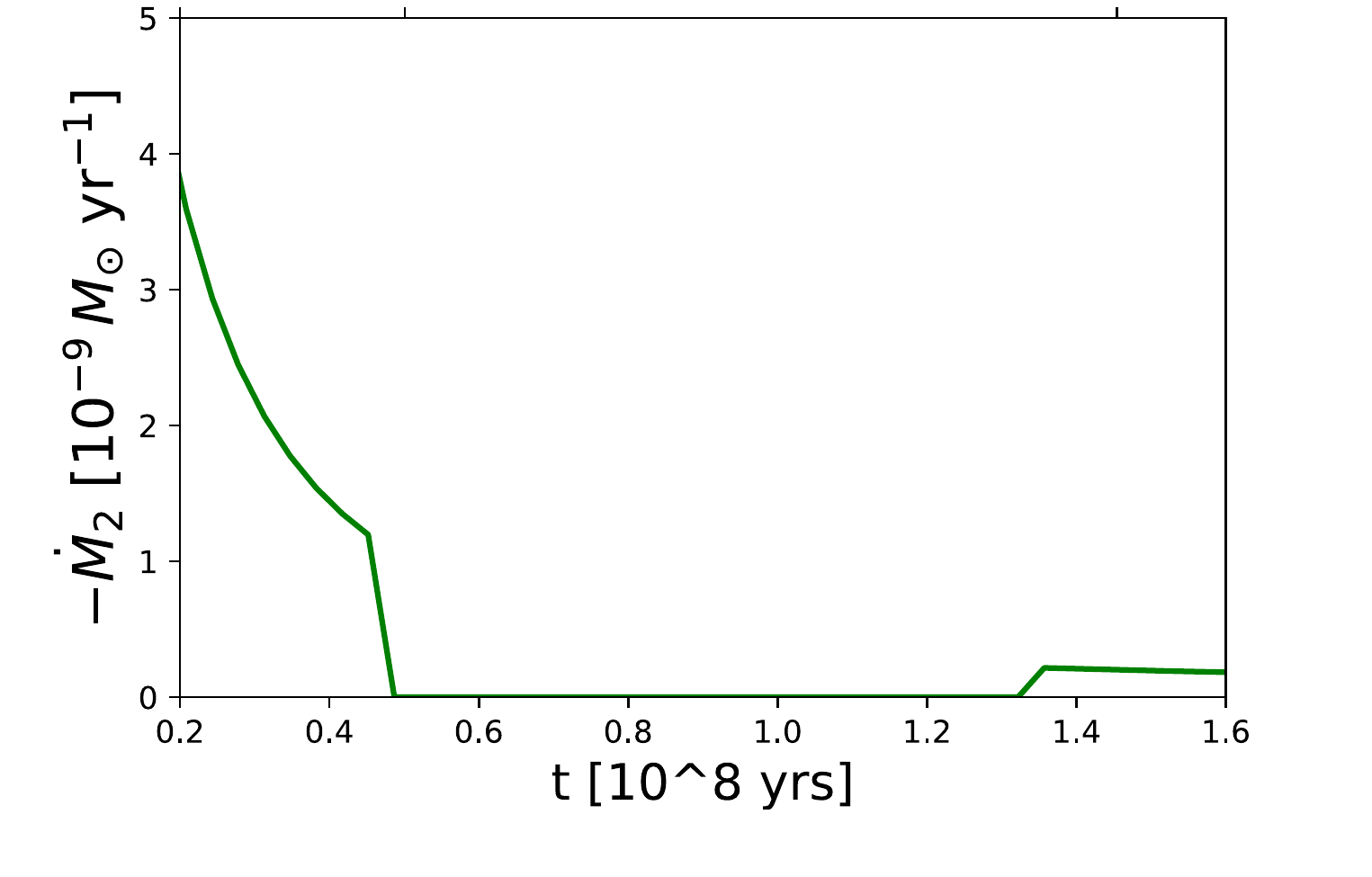}
 \caption{Single system evolution of the relative angular momentum $\dot{J}/J$ (top); the radius (green), the Roche-lobe (red), and the equilibrium radius (blue) of the secondary star (middle); and the corresponding mass accretion $\dot{M}$ (bottom). Note that because time increases towards the right and systems spin-up, orbital period increases towards the left, contrary to how is presented in Fig.~\ref{fig:mds}.}
\label{fig:mdot}
\end{figure}

Figure~\ref{fig:mdot} shows the angular momentum loss rate (top), the radius (middle), and the mass loss rate (bottom) evolution of a single CV system with time. We find a fast but smooth decrease ($\sim$ 90 \%) in magnetic braking near the upper boundary of the period gap that allows the secondary star to return to thermal equilibrium. As a consequence its radius (green line in bottom panel) decreases, detaching from its Roche lobe (red line in bottom panel) and mass accretion stops, as predicted by \cite{Spruit.Ritter:83}. 

{}

\section{Population Synthesis}
\label{ps}

We then synthesize a population of CVs to compare with histograms of the frequency of observed  CVs as a function of orbital period.  We use the same prescription that we used for a single system and that accounts for the complexity of the magnetic field in the angular momentum loss efficiency.  
Our primordial binary systems follow the usual assumptions.  We start from a  Zero Age CV (ZACV) population with a primary mass, $M_1$, from the \cite{Miller.Scalo:79} initial mass function,
$$ M_1(x) = 0.19x[(1-x)^{3/4}+0.032(1-x)^{1/4}]^{-1}$$
with masses in the range $0.8 < M_1 <8 M_{\odot}$ \citep{Howell.etal:01}, and a secondary mass from a probability distribution as in \cite{Abt.Levy:78, Halbwachs:87, Howell.etal:01},
\textbf{$f(q) = 5/4\, \, q^{1/4} $}, where $q=M_2/M_1$.
%and a uniform initial distribution over periods of 1 day to $10^6$ yr with circular orbits \citep[see, e. g., ][]{Abt.Levy:78, Duquennoy.Mayor:91}. 
In order to select only systems that undergo a common envelope phase we require that the radius of the Roche lobe
of the primary be larger than the radius of a star of mass $M_1$ at the base of the giant branch \citep[see, e.g., ][and references therein]{Paczynski:65, Webbink:79, Webbink:85, Webbink:92, deKool:92, Howell.etal:01}. We assume that a common envelope phase occurs and that the duration of the
spiral-in is sufficiently short ($10^4$ yr; see references above)
that the mass of the secondary does not change significantly
during the episode
\citep[see, e.g.,][]{Taam.etal:78, Meyer.Meyer-Hofmeister:79, Livio.Soker:88, Webbink:92,Rappaport.etal:94,Taam.Sandquist:98}. We then compute the final mass for the WD, $M_1=M_{core}$, as in \cite{Howell.etal:01}. 

Once we have generated the initial ZACVs masses, we compute the initial radius of the secondary assuming it is in thermal equilibrium, $R_2 = 0.9 M^{0.8}$, as we did in Section~\ref{aml}. This is justified since CVs spend most of their time in this regime and, if magnetic braking is fast enough, the system will quickly transition to a different $M_2$-$R_2$ regime.
We then calculate their orbital period using Equation~\ref{period} as for the single system evolution.
%We compute their final orbital separation based on energetic considerations \citep[see, e.\ g., ][]{Taam.etal:78, Meyer.Meyer-Hofmeister:79, Livio.Soker:88, Webbink:92,Rappaport.etal:94, Taam.Sandquist:98,  Howell.etal:00}.  
Out of an initial population of $\sim 10^6$, we typically end up with $10^4$ zero-age pre-CVs \citep[consistent with][]{Howell.etal:00}.   We generate a new set of ZACVs every $10^5$ years to account for a uniform rate of star formation.  
%The magnetic field complexity of each star is recalculated at each step as a function of rotation period. 

\begin{figure}{}
\center
   \includegraphics[trim = 1.3in 3.5in
  1.2in 3.5in,clip, width =  0.45\textwidth]{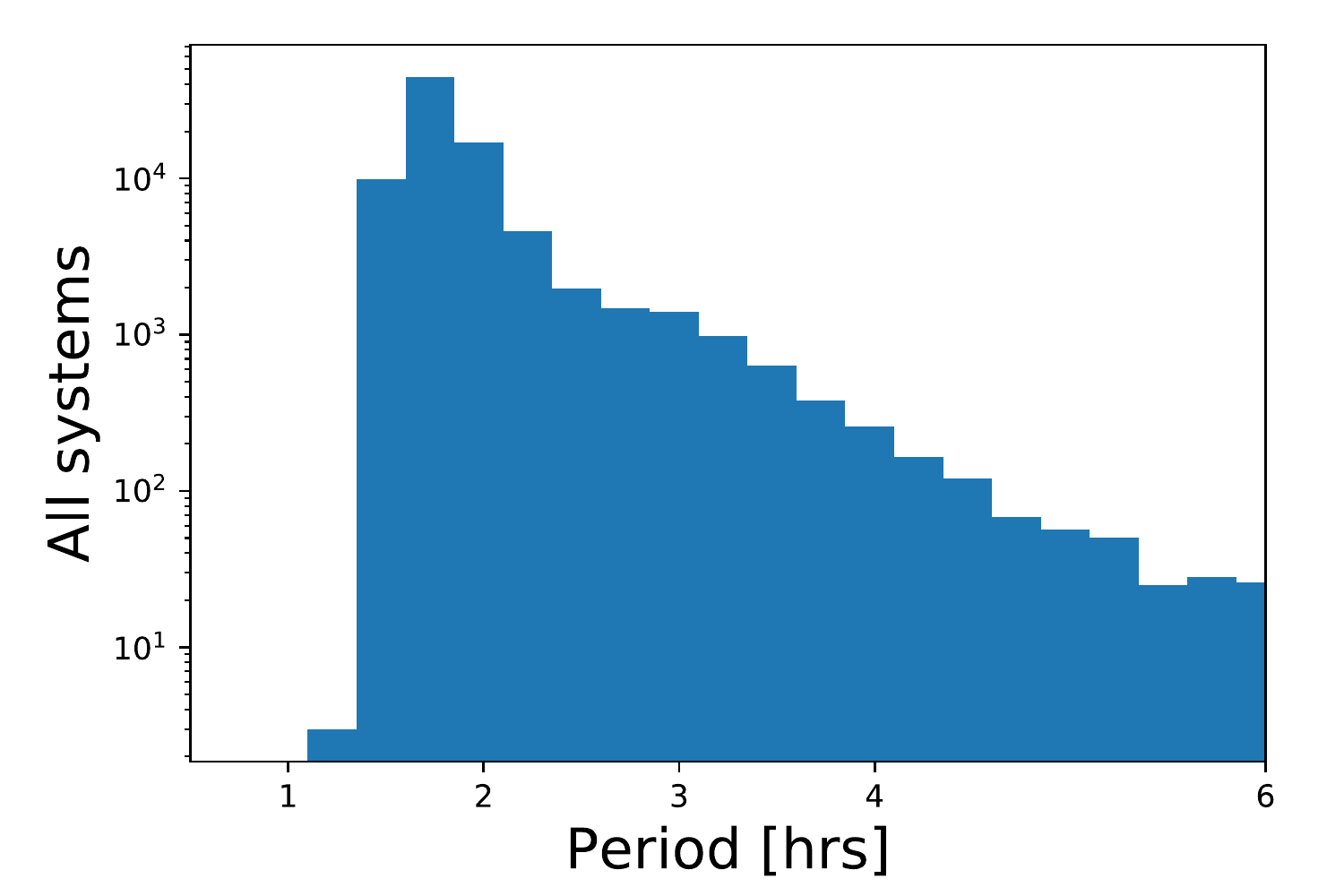} 
  \includegraphics[trim = 1.3in 3.5in
  1.2in 3.5in,clip, width =  0.45\textwidth]{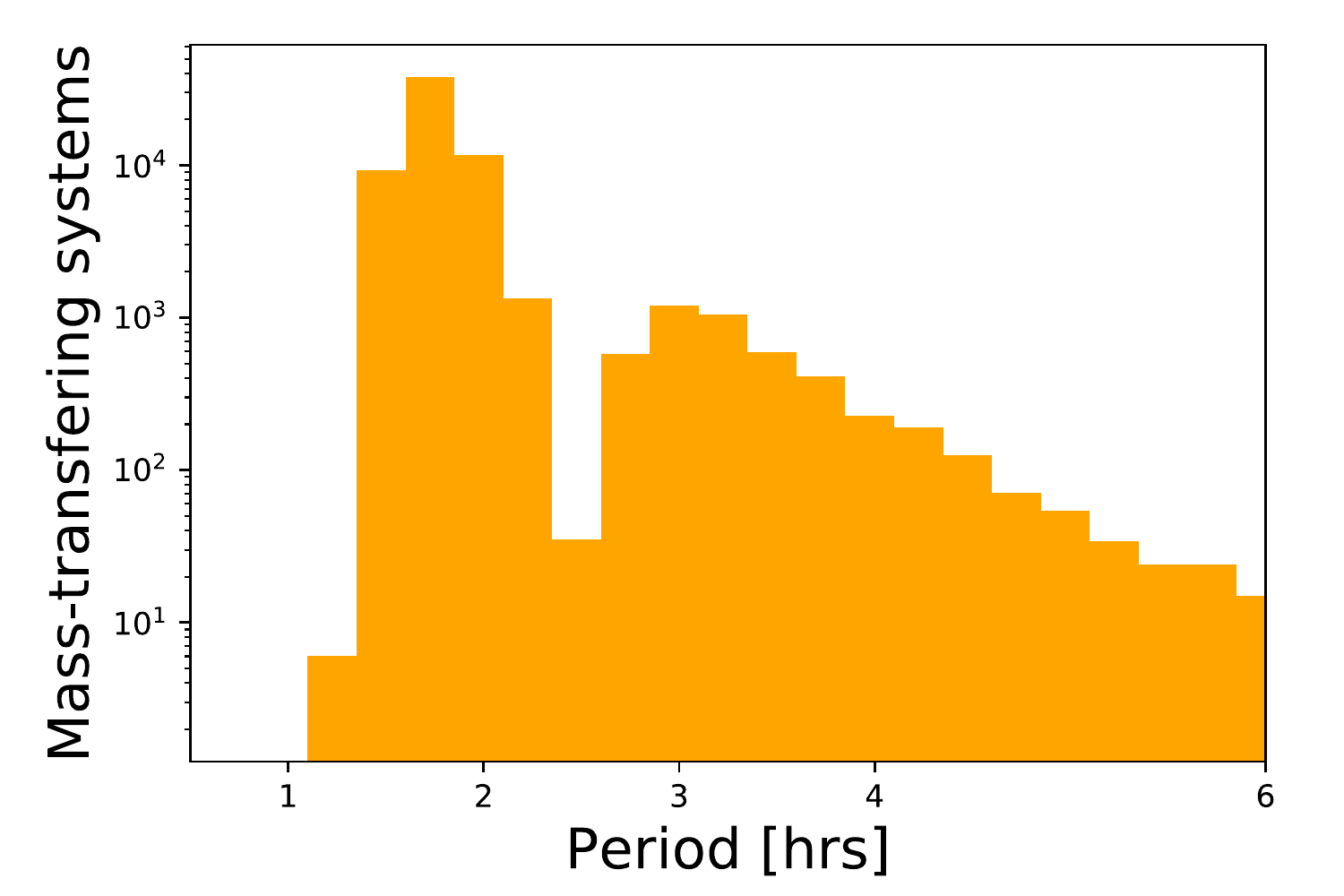} 
  \includegraphics[trim = 1.3in 3.5in
  1.2in 3.5in,clip, width =  0.45 \textwidth]{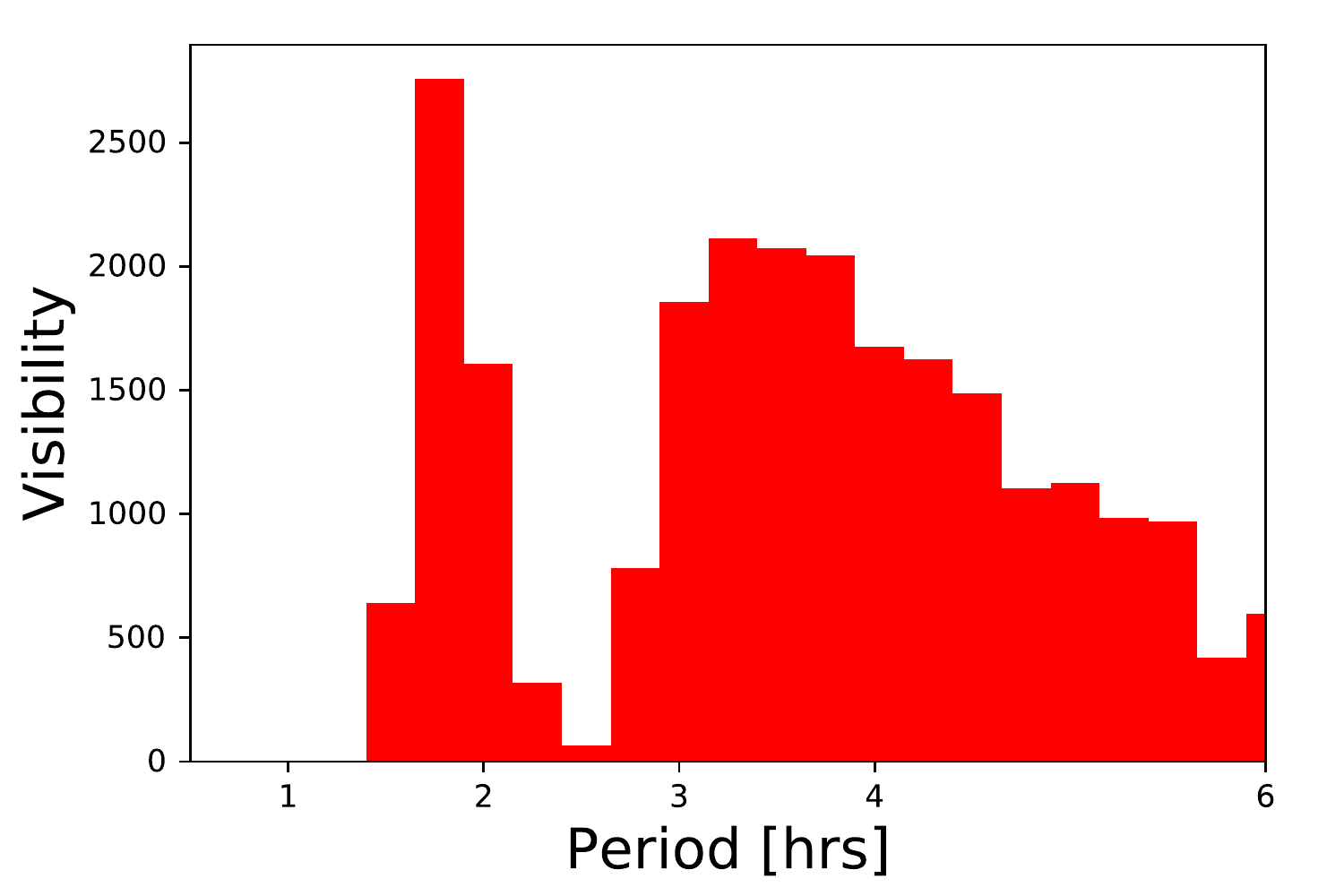} 
\caption{Histogram of expected number of systems (both accreting and not accreting) as a function of orbital periods (top), histogram of expected accreting systems (middle), and the sum of the accretion luminosity per bin in units of $10^{32}$ erg/sec (bottom). We find that 38\% of the systems lie above, 51\% below, and 11\% within the gap} %{\color{blue}JDAG: Since the period Gap is defined with such precision (up to the second decimal point in 3.12), would it make sense to make a finer binning for the histogram?} \textcolor{green}{CG: I am not sure about that since we don't have an agreement to such extent}}
\label{fig:visibility}
\end{figure}

%We know  from \cite{Spruit.Ritter:83} that an increase in complexity leading to $>90\%$ reduction in angular momentum loss efficiency will explain the period gap. Now we wish to find the complexity evolution as a function of period that will best reproduce it. 
We evolve these populations for $10^9$ years and produce a histogram for the number of systems in each period bin (see top panel of Figure~\ref{fig:visibility}). In addition, we make a histogram of the number of systems times their accretion luminosity, $L_{acc} \propto G M_1\dot{M_2}/R_1$ \citep[see, for example,][and references therein]{Knigge.etal:11}, which acts as a proxy for the observability for these systems (bottom panel, Figure~\ref{fig:visibility}).  
We find that as systems spin-up they accumulate at shorter periods, reflecting the decreasing magnetic braking (see Figure~\ref{fig:mdot}). The brightness of systems at periods of approximately 3 hours drops sharply and the period gap naturally arises from the magnetic disruption of angular momentum loss efficiency. They become visible again after they reach $P \approx 2$ hours, when spin-up rates increase again as a result of gravitational radiation becoming important. 

%acaaaaaaaaaaaaaa````````''''''''

The goal of this study is to show that stellar magnetic complexity evolution provides a natural explanation for the CV period gap while not requiring any change in the dynamo generation rate of magnetic flux, consistent with X-ray observations.  Explaining the period distribution of systems below the period gap is out of the scope of this paper. It is in that spirit that we use a simplified model for the period bouncers decrease in the mass-radius relation at short periods \citep{Knigge.etal:11}. 
%\begin{equation}
%n = \frac{0.02}{Ro}+Ro+1
%%n_{av} = 4.22e^{4}/(P-2.5*3600.)^{1.2} +1.
%\label{eq:nav}
%\end{equation}

{}

\section{Conclusions}
\label{results}

%\textcolor{blue}{
We have found that existing ZDI observations of late M dwarfs support a picture of increasing surface magnetic field complexity with decreasing rotation period for values of the period of several hours. This is consistent with the stellar spin-down model presented by CG18 that explains the bimodal rotation period distributions of stars in young open clusters in terms of evolving surface magnetic complexity. Greater magnetic complexity leads to suppression of mass loss and a shortening of the magnetic ``lever" that acts as the rotational brake in late-type stars. Consequently, as CVs evolve toward shorter periods they experience a reduction in angular momentum loss rate and a reduction in the accretion rate driven by magnetic braking, as conjectured by \citet{Taam.Spruit:89}.

%\textcolor{blue}{
We have modeled a synthetic population of CVs using the standard CV evolution equations \citep{Spruit.Ritter:83,Knigge.etal:11} together with the magnetic braking prescription provided in CG18.
As periods approach 3.2 hours, the reduction in angular momentum loss is so rapid and efficient ($\sim 90 \%$) that the accretion rate in most systems drops sufficiently to allow the puffed-up donor star to shrink back into thermal equilibrium.  These are just the conditions that \citet{Spruit.Ritter:83} pointed out would produce the CV period gap and that \citet{Taam.Spruit:89} found could arise from an increase in surface magnetic complexity. The secondary star no longer fills its Roche lobe and mass accretion stops, rendering it observationally inconspicuous. However, the system continues to lose angular momentum through magnetic braking at this slower rate and eventually (at $P \approx 2$ hours) the orbital separation decreases enough for accretion to resume and consequently for the system to become conspicuous again. Our model predicts the presence of a few systems accreting within the gap, consistent with observations.

%\textcolor{blue}{
The explanation of the period gap in terms of an increase in magnetic complexity of the secondary as systems approach the gap as first suggested by \citet{Taam.Spruit:89}, rather than the dynamo itself shutting down, is fully consistent with X-ray observations indicating that magnetic field generation is equally efficient above and below the M dwarf fully-convective limit. The disrupted magnetic braking idea of \citet{Spruit.Ritter:83} and \citet{Rappaport.etal:83} is not broken.

\acknowledgments

We thank Rosanne Di Stefano, Christian Knigge, Gerrit Schellenberger and Henk Spruit for useful comments and discussion, and the referee for helpful suggestions.
CG was supported by {\it Chandra} grant GO7-18017X, {\it XMM-Newton} grant 80NSSC18K0401 and HST grant GO-13754.0001-A during the course of this work. JJD was supported by NASA contract NAS8-03060 to the {\it Chandra X-ray Center}. JDAG was supported by Chandra grants AR4-15000X and GO5-16021X.  SPM was supported by NASA Living with a Star grant number NNX16AC11G. OC was supported by NASA Astrobiology Institute grant NNX15AE05G. Simulation results were obtained using the Space Weather Modeling Framework, developed by the Center for Space Environment Modeling, at the University of Michigan with funding support from NASA ESS, NASA ESTO-CT, NSF KDI, and DoD MURI. Simulations were performed on NASA's PLEIADES cluster under award SMD-16-6857.  

%\bibliographystyle{apj}
%\bibliography{AML} 

\end{document}